
\input jnl

\preprintno{ILL--(TH)--92--7}

\title{Three--Dimensional Quantum Gravity Coupled to Ising Matter}

\author{Ray L. Renken}

\affil
Department of Physics
University of Central Florida
Orlando, Florida  32816

\author{Simon M. Catterall}

\affil
Department of Applied Math and Theoretical Physics
University of Cambridge
Silver St., Cambridge  CB3 9EW
England

\author{John B. Kogut}

\affil
Loomis Laboratory of Physics
University of Illinois at Urbana--Champaign
1110 West Green Street
Urbana, Illinois  61801

\abstract{We establish the phase diagram of three--dimensional quantum
gravity coupled to Ising matter.  We find that in the negative curvature
phase of the quantum gravity there is no disordered phase for
ferromagnetic Ising matter because the coordination number of the sites
diverges.  In the positive curvature phase of the quantum gravity there is
evidence for two spin phases with a first order transition between them. }

One of the results that propelled recent work on two dimensional quantum
gravity was the discovery that dynamically triangulated random surfaces
provide a convenient formulation of the theory and facilitate both analytic
and numerical calculations.  An analogous formulation of three dimensional
quantum gravity has also been proposed [1] with some success [2].  This
success is non--trivial because the proof for two dimensions that the number
of triangulations is exponentially bounded as a function of the number of
triangles does not carry over to three dimensions.  In two dimensions, the
classification of topologies in terms of genus is crucial in proving the bound.
In three dimensions, there is no comparable topological classification scheme.
Without the exponential bound, the naive result is to have factorial growth.
This would mean that no chemical potential term could control the theory.
However, numerical experiments suggest that there is an exponential bound,
at least when the topology is restricted to $S^3$ [3].

Another difference in three dimensions is the need for a second chemical
potential.  In two dimensions, the Euler characteristic, $\chi$, is
$$N_0 - N_1 + N_2 = \chi\eqno(1)$$
(where $N_i$ is the number of i--simplices). Furthermore, each triangle
has three
links and each link is shared by two triangles implying
$$3N_2 = 2N_1\eqno(2)$$
The most general action of the form
$$S = \sum g_i N_i\eqno(3)$$
therefore has only one independent term.
This can be taken to be a cosmological constant determining the mean number
of triangles. In two dimensions it
is further possible to restrict ourselves to a ``microcanonical'' ensemble
which eliminates even this term.  The corresponding equations in three
dimensions are
(for $\chi = 0$, which will be assumed from here on)
$$N_0 - N_1 + N_2 - N_3 = 0\eqno(4)$$
and
$$4N_3 = 2N_2\eqno(5)$$
leaving two independent terms in an action of the form (3).
This
action can be chosen to be
$$S = \alpha N_0 - \beta N_3\eqno(6)$$
so that $\beta$ controls the number of tetrahedra (the volume) and $\alpha$
controls the number of nodes.  In the continuum, $\beta$ corresponds to the
cosmological constant while $\alpha$ is related to
the gravitational constant and hence the mean curvature.
Unfortunately, there are no proofs of ergodicity for
a microcanonical ensemble in three dimensions, so we are forced to
utilize a canonical ensemble for our simulations. Of course, the
existence of a continuum limit is a major issue.  The question is whether
or not there is a second order phase transition in the $(\alpha,\beta)$ plane
(or in some larger space if actions of a more general form than (3) are
considered).  The coupling $\beta$ must be adjusted so that the volume
approaches
infinity.  This leaves $\alpha$ for our attention.  It is reported that for
$\alpha < \alpha^*$ (where $\alpha^* \sim 3.75-4.0$), the theory is in a
negative curvature phase while for $\alpha > \alpha^*$ the mean
curvature is positive [4].
Simulations suggest that the transition between them
may be first order [5] implying that no continuum limit is available.
It has been speculated that this negative result is related to
the topological nature of pure three dimensional gravity in which there
is no massless graviton.

Matter can be coupled to the gravity by placing it at the
nodes (for instance) and adding an appropriate term to the action.  In two
dimensions, there is some understanding of the effects of coupling quantum
gravity to matter, at least when the central charge is less than one.
Knihznik, Polyakov, and Zamalodchikov (KPZ) [6] have argued that when the
central charge is less than one, the scaling dimensions of the matter fields
are
simply ``dressed'' by quantum gravity.  On the lattice this means that for
a critical point of some matter theory on a fixed lattice, there is a
corresponding second order phase transition for the matter on a dynamical
lattice, but with different critical exponents.  KPZ give an explicit relation
for the two sets of scaling dimensions.  In particular, the Ising model
has a continuous phase transition when coupled to quantum gravity with
exponents different from the usual ones.
In this paper, we couple the Ising model to three dimensional quantum
gravity [7].
It is possible that within this enlarged parameter space we
may be able to find new critical regions where a
continuum limit could be taken. There is no analog of the KPZ framework
to guide us, so, clearly, a central
issue is whether or not an Ising transition continues to exist and, if so,
what its order is.

We introduce the matter by simply decorating the nodes of each diagram with
spins, adding an Ising term to the action with coupling $k$, and tracing over
spins $t_i=\pm 1$.  Thus our total
partition function $Z$ is given by
$$Z\left(\alpha,\beta,k\right)=\sum_{\rm T,\chi=0}\sum_{t_i}
 \exp{\left(\alpha N_0-\beta N_3+k\sum_{<ij>}t_it_j\right)}\eqno(7)$$

There are two important relations between the partition function of
this new theory and the partition function of pure gravity.  The first is
due to the fact that the trace over spins contributes a factor $2^{N_0}$
even if the Ising coupling is zero:
$$Z(\alpha,\beta,k=0) = Z_{pure~grav}(\alpha + \ln 2,\beta)\eqno(8)$$
In other words, in a grand canonical ensemble, the Ising spins
renormalize the gravitational couplings even at
zero coupling. For the
new action, this implies that the curvature transition is in the range
$\alpha^* \sim 3.06-3.31$.
The second relation is due to the fact that once the spins are completely
ordered, the Ising term just contributes a factor $k N_1$ to the action.
$$\null\qquad Z(\alpha,\beta,k) \sim Z_{pure~grav}(\alpha+k,\beta-k)
\qquad {\rm (for~k~such~that~}<tt^\prime>\sim N_1)\eqno(9)$$
One implication of this last result is that starting with any $\alpha$ in the
negative curvature phase, large enough $k$ will take the system into the
positive curvature phase.
In the discussion that follows we always assume that $\beta$ has been adjusted
to fix the mean volume associated with the tetrahedra. The
continuum limit will finally be taken by letting this mean value
tend to infinity. This leaves us
with a two dimensional phase diagram.  (This can be done because the empirical
exponential bound for the pure gravity implies an exponential bound in the
presence of matter).

Before proceeding with a numerical analysis of the phase diagram it is useful
to consider the Ising model in mean field theory on {\it quenched} lattices
of the type found in the pure gravity theory.  This will give a kind of
zeroth order approximation to the phase diagram and will help direct the
numerical work.

Take $N_0$ and $N_3$ to be fixed and make the standard mean field approximation
$$         \sum tt^\prime  = q <t> \sum t \eqno(10)$$
where $tt^\prime$ is the Ising term, $<t>$ is a spontaneous magnetization,
and $q$
is the average coordination number. As usual we may add an external
magnetic field to facilitate the computation of
expectation values.  The partition function now reads
$$         Z = {\rm (const)~Tr~}\exp([h + kq<t>] \sum t) \eqno(11)$$
where $h$ is the magnetic field and $k$ is the Ising coupling.  This is now a
non-interacting system so the trace can be done to get
$$         Z = {\rm (const)~(2~}\cosh[h+kq<t>])^{N_0}\eqno(12)$$
The average spin can be calculated from this and is
$$        <t> = {1\over N_0} {d \ln Z \over  d h} = \tanh(h+kq<t>)\eqno(13)$$
For $h=0$ and small $<t>$ (i.e. near the critical point) $<t> = kq<t>$, so the
critical value of the Ising coupling is $$k = 1/q.\eqno(14)$$

Equations (10--14) are standard.  What is new is the effect of the
background lattice via the coordination number $q$.  Each link contributes
to the coordination number of two sites, so
$$2 N_1 = N_0 q\eqno(15)$$
Equations (14) and (15), together with $N_1 = N_0 + N_3$ (from (4) and (5)),
give
$$k = {1\over 2}{1\over (1+(N_3/N_0))}\eqno(16)$$
Now the problem boils down to the ratio of $N_3/N_0$ as $N_3$ goes to infinity.
Ref. [4] finds that for large $N_3$,  $<N_0> = N_3^{(\delta)}$ with $\delta <
1$
in the negative curvature phase and $\delta = 1$ in the positive curvature
phase.  When $\delta < 1$, $N_3/N_0$ goes to infinity
(which just means the average coordination number goes to infinity) and
the critical
coupling is at zero.  This is what happens at all points in the negative
curvature phase.  When $\delta = 1$, i.e. $<N_0> = N_3 / c$, we get
$$k = {1\over 2}{1\over (1+c)}\eqno(17)$$
Ref. [4] gets (in pure gravity) for $\alpha=4$, $c = 3.368$.  This gives
$k = 0.114$ (which turns out to be very close to our numerical result).

This mean field argument suggests the phase diagram of fig. 1.  The dashed
line is just eqn. (9) with the knowledge that the negative curvature
phase orders the spins for arbitrarily small couplings.  This line then
separates
the negative curvature phase from the positive curvature phase.  The thick
line is equations (16) and (17) with input about the relationship
between $N_0$ and $N_3$ in the two phases.  Since mean field theory is an
expansion in $1/q$, when $q = \infty$ it should be exact.  In the negative
curvature phase, therefore, the Ising spins order for arbitrarily small
positive values of the Ising coupling.  The only part of the phase diagram
in question is the positive curvature phase part of the Ising line.  We
address this question numerically.

The gravity sector of the theory is simulated using update moves that have
been shown to be ergodic [8] and have become standard.  Barycentric
subdivision,
also called the (1,4) move, takes a tetrahedra, inserts a node in the middle
and connects this node to the other four vertices creating a total of four
tetrahedra.  The analog of the bond flip in two--dimensions, here the (2,3)
move, removes the triangle shared by two tetrahedra, introduces a link between
the two nodes that are not shared and places three new triangles around this
new link (one for each of the shared nodes) in such a way that there are now
three tetrahedra.

There is a fine tuning problem associated with this theory.  If the number
of simplicial manifolds regarded as a function of the number of tetrahedra
behaves as
$$\#(N_3) \sim (N_3)^a e^{c N_3}\eqno(18)$$
with $a$ negative (such as happens for two dimensions), then the
grand canonical partition function is dominated by large volumes if
the chemical potential is slightly too small and by small volumes
if the chemical potential is slightly too large.  In
numerical simulations this manifests itself
as the tendency for the system to evolve to one of the two extremes
no matter how carefully the couplings are adjusted.
Several approaches to this
difficulty have been used.  One group confined their simulations to a range of
$N_3$ [4].  We prefer to retain an action formulation of the problem and add a
gaussian term to the action as proposed in [2] (except that we do not constrain
$N_0$):
$$\delta S = - \gamma (N_3 - M_3)^2\eqno(19)$$
The desired volume $M_3$ is put in by hand.  The coupling $\beta$ is then
adjusted so that $<N_3> = M_3$.  The new coupling $\gamma$ is made small so
that it does not influence the physics of interest.  This new term in the
action effectively ensures that the probability
distribution for $N_3$ is approximately  gaussian about the mean value
$M_3$, with a
characteristic width determined by $\gamma$. The length of the simulation
is then tuned with
$\gamma$ so that the system has time to wander
back and forth across the distribution many times.  Otherwise,
the system may not
reach equilibrium.  In practice, $\gamma = 0.005$ is a practical value.

With the moves and action now defined, it is important that
detailed balance must be built into the update probabilities [9].
For pure gravity, two lists are made, one with all sets of two or three
tetrahedra for which the 23 or 32 move is allowed and another with all sets of
one or four tetrahedra for which the 14 or 41 move is allowed.  Each list
consists of items in one of two possible states which we will call A and B.
There is some number, $n$, of items in the list.  Picking a state at random and
updating it produces a modified list with $n^{\prime}$ items.  Detailed balance
is maintained if the updating probability is
$$P(A\rightarrow B)
 = {1 \over 1 + {n^{\prime} \over n} e^{S_A - S_B} }\eqno(20)$$
Adding the Ising matter requires some modifications.  Except for its effect on
the action,
no modification is necessary for the 23 or 32 moves.  However, the update
probabilities must be modified for the 14 and 41 moves because the list now
has three possible states (no node, node spin up, node spin down) rather than
two.  Call the spin states $B1$ and $B2$.  A suitable modification of the
update probabilities is
$$P(A\rightarrow B1) = {n e^{S_{B1}} \over n^{\prime} e^{S_A} +
n (e^{S_{B1}} + e^{S_{B2}}) }$$
$$P(B1\rightarrow A) = {n^{\prime} e^{S_A} \over n^{\prime} e^{S_A} +
n (e^{S_{B1}} + e^{S_{B2}}) }\eqno(21)$$
$$P(B1\rightarrow B2) = {n e^{S_{B2}} \over n^{\prime} e^{S_A} +
n (e^{S_{B1}} + e^{S_{B2}}) }$$
and three more expressions with 1 and 2 interchanged.  It is most convenient
to consider the total probability of introducing a new node,
$P(A\rightarrow B1) + P(A\rightarrow B2)$, and to determine the spin value
(using Metropolis) after deciding whether or not to put in the new node.

Of course, there are other ways to maintain detailed balance.  One can choose
from lists of moves which are not necessarily allowed and then test for this
during the accept/reject step.  In the approach above, it is most time
consuming to predict the number of elements that will be in the list of
allowed (23) moves if a contemplated (23) move is made.  It is more efficient
to postpone part of this calculation to the accept/reject step (the part that
checks whether or not a candidate simplex already exists or not) while keeping
the simpler restrictions on the list (such as requiring links to have the
correct coordination number).

Calculations were performed for values of $\alpha$ in the range 3.5 to 4.0
and for values of the Ising coupling between 0 and 0.2.  As $k$ is varied in
this range the nearest neighbor spin expectation value varies from 0 to
around 0.7 (i.e. from complete disorder to fairly ordered).  The Ising
component of the specific heat was used
to search for a possible spin ordering  transition and the Ferrenberg--Swendsen
approach was used to cover a range of $k$ (and sometimes $\alpha$) from a
single simulation.  At $\alpha = 3.5$, on small lattices (up to $N_3 = 1600$),
no peak was observed in the specific heat as a function of the Ising coupling
and correlation times remained small.
Instead, the specific heat was observed to rise to a plateau as $k$
was increased.  For large
$k$, nearest neighbor spins are strongly correlated and the fluctuation in
the number of nodes contributes significantly to the specific heat.
This is the origin of the plateau.  On larger lattices, $N_3 = 3200$ and
$N_3 = 6400$, a peak appears above the plateau value and grows (along with an
increased correlation time) suggesting the existence of a phase transition.
Figure 2 shows the peak just beginning to appear for $3200$ tetrahedra.

Very long runs (over half a million
sweeps) coupled with the Ferrenberg--Swensen method allowed us to calculate
the kurtosis ($1-<E^4>/3<E^2>^2$ where $E = k\sum tt$) of the distribution of
the spin correlation for couplings near that of the specific heat peak [11].
Tables 1 and 2 give a sample of the results.
The minimum value was near 0.48 for both volumes while the value at the peak
of the specific heat increased with increasing volume.  The most naive
extrapolation to infinite volume still leaves the kurtosis significantly
below its gaussian value (2/3) indicating a first order transition.
The peak of the specific heat itself roughly doubles as the volume is doubled,
another hint that the transition is first order.

What happens as $\alpha$ is increased?  For significantly greater values of
$\alpha$ we were unable to find a peak rising above the plateau.  It is
possible that the first order transition ends and is replaced with a smooth
crossover behavior, but it appears more likely that it is simply necessary
to go to larger lattices in order to continue to see the transition.

In conclusion, the ferromagnetic sector of the phase diagram of
three dimensional quantum gravity coupled to Ising spins appears to have
three components (see figure 1).  Phase 1 has ordered Ising spins and negative
curvature.  Phase 2 has ordered Ising spins and positive curvature.  Phase 3
has disordered Ising spins and positive curvature.  Near the gravitational
transition, the transition from phase 2 to phase 3 appears to be first order.
For larger $\alpha$ the nature of the transition is uncertain. Further
calculations on larger lattices are needed to resolve this question
and to verify the order of the transition.

This work was supported, in part, by NSF grant PHY 87--01775.  Some
calculations
were performed on the Florida State University Cray Y--MP and at the National
Center for Supercomputer Applications in Illinois.  Also, we acknowledge
National Science Foundation Support through the Materials Research Laboratory
at the University of Illinois, grant NSF--DMR--20538.
\references

[1]  N. Godfrey and M. Gross, \prd 43, R1749, 1991.

[2]  M. E. Agishtein and A. A. Migdal, \journal Mod. Phys. Lett., A6, 1863,
1991 and PUPT--1272.

[3]  J. Ambjorn and S. Varsted, NBI--HE--91--17.

[4]  D. V. Boulatov and A. Krzywicki, LPTHE Orsay 91/35.

[5]  J. Ambjorn and S. Varsted, NBI--HE--91--45,6.

[6]  V.G. Knizhnik, A.M. Polyakov and A.B. Zamolodchikov, \journal
Mod. Phys. Lett., A3, 819, 1988.

[7]  A different approach is considered in C. F. Baillie,
COLO--HEP--279.

[8]  M. Gross and S. Varsted, NBI--HE--91--33.

[9]  We thank Mark Gross for very helpful discussions.

[10]  A. M. Ferrenberg and R. H. Swendsen, \journal Phys. Rev. Lett., 23,
2635, 1988.

[11]  M.S.S. Challa, D.P. Landau, and K. Binder, \journal Phys. Rev., B34,
1841,1986.
2635, 1988.

\endreferences

\figurecaptions
[1]  The phase diagram suggested by quenched mean field theory.

[2]  The specific heat, $C$, versus the Ising coupling for $3200$
tetrahedra.

\endfigurecaptions

\setbox\strutbox=\hbox{\vrule height11.5pt depth5.5pt width0pt}
$$\vbox{\tabskip=0pt \offinterlineskip
\halign to200pt{\strut#& \vrule#\tabskip=1em plus2em& \hfil#& \vrule#&
\hfil#\hfil& \vrule#& \hfil#& \vrule#
\tabskip=0pt\cr \noalign{\hrule}
& & \omit\hidewidth $k$\hidewidth
& & \omit\hidewidth $C$\hidewidth
& & \omit\hidewidth $\kappa$\hidewidth
& \cr \noalign{\hrule}
&& 0.110 && 0.125 && 0.653 & \cr\noalign{\hrule}
&& 0.115 && 0.135 && 0.566 & \cr\noalign{\hrule}
&& 0.118 && 0.136 && 0.495 & \cr\noalign{\hrule}
&& 0.120 && 0.136 && 0.479 & \cr\noalign{\hrule}
&& 0.122 && 0.135 && 0.495 & \cr\noalign{\hrule}
&& 0.125 && 0.132 && 0.548 & \cr\noalign{\hrule}
&& 0.129 && 0.125 && 0.612 & \cr\noalign{\hrule}
}}$$
Table 1.  Ferrenberg--Swendsen estimates of the specific heat, $C$, and
kurtosis, $\kappa$, as a function of Ising coupling for $3200$ tetrahedra.
The actual coupling for the run was 0.120.

\setbox\strutbox=\hbox{\vrule height11.5pt depth5.5pt width0pt}
$$\vbox{\tabskip=0pt \offinterlineskip
\halign to200pt{\strut#& \vrule#\tabskip=1em plus2em& \hfil#& \vrule#&
\hfil#\hfil& \vrule#& \hfil#& \vrule#
\tabskip=0pt\cr \noalign{\hrule}
& & \omit\hidewidth $k$\hidewidth
& & \omit\hidewidth $C$\hidewidth
& & \omit\hidewidth $\kappa$\hidewidth
& \cr \noalign{\hrule}
&& 0.0790 && 0.029 && 0.653 & \cr\noalign{\hrule}
&& 0.0820 && 0.062 && 0.568 & \cr\noalign{\hrule}
&& 0.0844 && 0.154 && 0.474 & \cr\noalign{\hrule}
&& 0.0866 && 0.248 && 0.557 & \cr\noalign{\hrule}
&& 0.0900 && 0.124 && 0.613 & \cr\noalign{\hrule}
&& 0.0920 && 0.092 && 0.608 & \cr\noalign{\hrule}
&& 0.0960 && 0.079 && 0.631 & \cr\noalign{\hrule}
}}$$
Table 2.  Ferrenberg--Swendsen estimates of the specific heat, $C$, and
kurtosis, $\kappa$, as a function of Ising coupling for $6400$ tetrahedra.
The actual coupling for the run was 0.090.
\endit